\documentstyle[12pt,epsf]{article}

\def\hybrid{
      \topmargin-20pt      
      \oddsidemargin 0pt
      \headheight 0pt   \headsep 0pt
     \textwidth 6.5in  
     \textheight 9in         
      \textwidth 6.25in 
      \textheight 9.5in 
      \marginparwidth .875in
      \parskip 5pt plus 1pt   \jot = 1.5ex}
\hybrid

\def\x{\times}

\def\o+{\oplus}
\def\ra{\rightarrow}
\def\beqa{\begin{eqnarray}}
\def\eeqa{\end{eqnarray}}

\newcommand{\ps}{\phantom{-}}
     
\newcommand\figinsert[4]
{\begin{figure}[htb]
\newcommand\figsize{#3}
\epsfysize\figsize
\vskip +.5cm
\centerline{\epsffile{#4}}
\vskip +.5cm
\caption{\leftskip 1pc \rightskip 1pc
\baselineskip=10pt plus 2pt minus 0pt {\sl {#2}}}
\label{fig:#1}
\vskip -.3cm\end{figure}}

\begin{document}
\thispagestyle{empty}
\rightline{IFP-9902-UNC}
\rightline{IASSNS-HEP-99-04}
\rightline{hep-th/9903052}
\vspace{2truecm}
\centerline{\bf \LARGE Towards the Standard Model spectrum}
\vspace{0.3truecm}
\centerline{\bf \LARGE from elliptic Calabi-Yau}

\vspace{1.5truecm}
\centerline{Bj\"orn Andreas$^{\dagger}$
\footnote{bandreas@physics.unc.edu, 
supported by U.S. DOE grant DE-FG05-85ER40219/Task A.}, 
Gottfried Curio$^*$\footnote{curio@ias.edu, 
supported by NSF grant DMS9627351} 
and Albrecht Klemm$^*$\footnote{klemm@ias.edu,  supported by Heisenberg 
Fellowship 
and NSF grant DMS9627351}}
\vspace{.6truecm}

{\em
\centerline{$^{\dagger}$Department of Physics and Astronomy}
\centerline{University of North Carolina at Chapel Hill, 
NC 27599-3255, USA}
\vspace{.4truecm}
\centerline{and}
\vspace{.4truecm}
\centerline{$^*$School of Natural Sciences, IAS}
\centerline{Olden Lane, Princeton, NJ 08540, USA}}

\vspace{1.0truecm}
\begin{abstract}
We show that it is possible to construct supersymmetric three-generation models 
of 
Standard Model gauge group in the framework of non-simply-connected
elliptically fibered Calabi-Yau, without section but with a bi-section. 
The fibrations on a cover Calabi-Yau, where the model has $6$ generations of 
$SU(5)$
and the bundle is given via the spectral cover description, use a different 
description of the elliptic fibre which leads to more than one global section. 
We present two examples of a possible cover Calabi-Yau with a free involution: 
one is a fibre product of rational elliptic surfaces $dP_9\, $; another example 
is 
an elliptic fibration over a Hirzebruch surface. There we give the necessary 
amount of 
chiral matter by turning on in the bundles a further parameter, related to 
singularities 
of the fibration and the branching of the spectral cover.
\end{abstract}

\bigskip \bigskip
\newpage

\section{Introduction}

Efforts to get a (supersymmetric) phenomenological spectrum 
{}From the $E_8\times E_8$ heterotic string on a Calabi-Yau $Z$
started with embedding the spin connection in the gauge connection 
which gave an unbroken $E_6$ (times a hidden $E_8$
which couples only gravitationally). 
This procedure was combined in a second step with a further breaking
of the gauge group by turning on Wilson lines using a non-trivial
$\pi_1 (Z)$. The simplest constructions of Calabi-Yau spaces, as 
hypersurfaces in toric varieties will always have a trivial 
fundamental group. However one can still produce $\pi=G$ by dividing 
$Z$ by a freely acting group $G$, provided such an operation on 
$Z$ exists. This then led at the same time to a reduction of the 
often high number of generations 
(being equal to $\chi(Z)/2$) by the order of $G$.

This approach was generalised [\ref{W}] to the case of embedding 
instead of the tangent bundle an $SU(n)$ bundle for $n=4$ or $5$,
leading to unbroken $SO(10)$ resp. $SU(5)$ of even greater 
phenomenological interest then $E_6$. This subject was revived when,
as a consequence of the investigation of heterotic/$F$-theory duality,
the bundle construction was made much more explicit for the case of
elliptically fibered Calabi-Yau $\pi: Z\ra B$ [\ref{FMW}]. This extended 
ansatz showed among other things to have a much greater flexibility
in providing one with three-generation models (of the corresponding
unbroken GUT group) [\ref{C}].

There remains to go to the Standard Model gauge group in this 
framework\footnote{Three generation models with Standard model 
gauge group were constructed in orbifold and free fermion models 
before but these are, in contrast to our models,
defined only at isolated points in moduli space.}. 
For the following bases one can show that the elliptic 
fibrations, we will discuss, are smooth: the Hirzebruch 
surfaces $F_m$, $m=0,1,2$, the del Pezzo surfaces $dP_k$ 
with $k=0, \, ...6$, the rational elliptic surface 
$dP_9$ and 10 additional examples from the list of classified 
[\ref{koelman}] toric varieties, which correspond to two dimensional 
reflexive polyhedra. 
However none\footnote{The Enriques surface as base $B$ 
has $\pi_1(B)={\bf Z}_2$ leading to a non-trivial $\pi_1(Z)$ 
but, as pointed out by E. Witten, 
does not lead to a three generation model (cf. [\ref{DLOW}]).} 
of them has non-trivial $\pi_1(Z)$. 

The purpose of this note is to show that the elliptic framework 
is nevertheless 
capable of providing one with a three generation model of Standard Model 
gauge group by giving just a typical example how to proceed. 
Namely\footnote{following
a suggestion of E. Witten} we will look for an elliptically fibered
Calabi-Yau $Z$ which has besides the usually assumed section 
of its elliptic fibration a second section (a construction which
we will call $B$ model in contrast to the usual form, here called 
$A$ model); this will lead to 
a free involution on $Z$ which after modding it out provides you with 
a {\em smooth} Calabi-Yau where one now has 
the possibility of turning on Wilson lines which break $SU(5)$ to the
Standard Model gauge group. Thereby we will have achieved our goal
when we specify on $Z$ an $SU(5)$ bundle that leads to $6$ generations
and fulfills some further requirements of the 
spectral cover construction. A variant of this procedure would be 
to consider a fibration with two ${\bf Z}_2$ operations ($D$ model)
and to break $SO(10)$ to the Standard model gauge group times 
$U(1)_{B-L}$; we will comment on this possibility\footnote{
We will have not use for a symmetry breaking pattern for $E_6$
as in this case one needs at least a ${\bf Z}_7$ [\ref{Wsy}]
which we do not get in our approach.} too.

In {\em section 2} the spaces along with their cohomological 
data are introduced. In {\em section 3} the spectral cover construction
of bundles is recalled and the necessary adjustments for our base 
Calabi-Yau as well as some consistency conditions are spelled out; 
finally, there is given an example of a $6$ generation model 
('above') for $B=F_1$. In  {\em section 4} the question of
modding these spaces/bundles is treated; for $F_0$, $F_2$ and a type of $dP_9$ 
(with a $B$-type elliptic fibre) 
we succeed in writing down a free involution. 
In {\em section 5} we point to a method to get
a new class
of bundles, which though is not capable of producing chiral matter.
Finally in {\em section 6} we
turn on a further possible parameter (an option 
already existent in the $A$ model), related to singularities of the 
fibration  and  the branching of the spectral cover, to
generate chiral matter; here we give for $B=F_0$ or $F_2$ for the first 
time moddable $6$ generation models.

\section{The spaces}
\label{spaces}

\subsection{Change of fibre type}

As it will be our goal to mod not just the Calabi-Yau spaces but 
also the geometric data describing the bundle (and this transformation
of bundle data into geometric data uses in an essential way the 
elliptic fibration structure) we will search only for operations which
preserve the fibration structure, i.e. $\underline{\iota}\cdot \pi=\pi\cdot 
\iota$ 
where the $\underline{\iota}$ is an action on the base.  
This has the consequence that our fibrations will have $n$ sections
for $n=|G|$.

To have this extra structure, which allows for free ${\bf Z}_n$ actions  
on the elliptically fibered Calabi-Yau we will use a different 
elliptic curve than the usually taken ${\bf P}_{1,2,3}(6)$, which we refer
to as the $A$-fibre. We will have to  extend the construction of 
[\ref{FMW}] to the  new fibre types, 
labeled by $B,C,D$

\begin{center}
\begin{tabular}{c|c|c}
  & $(y,x,z)\in$         & $E$ \\ \hline
A & ${\bf P}_{3,2,1}(6)$ & $y^2+x^3+z^6+sxz^4=0$ \\ \hline
B & ${\bf P}_{2,1,1}(4)$ & $y^2+x^4+z^4+sx^2z^2=0$ \\ \hline
C & ${\bf P}_{1,1,1}(3)$ & $x^3+y^3+z^3+sxyz=0$ \\ \hline
D & ${\bf P}^3(2,2)$     & $x^2+y^2+szw=0,\; z^2+w^2+sxy=0$
\end{tabular}\nobreak
\newline\nobreak
\vskip 2 mm \nobreak
{\bf Tab. 1} {\footnotesize \sl The $A,B,C,D$ fibre types}\nobreak
\end{center}

These different possibilities were considered  before, mainly in the 
context of $F$-theory [\ref{KLM}-\ref{KLRY}]. In particular [\ref{BKMT}] 
contains an in depth study of these fibrations over a 
{\em one-dimensional} base (i.e. ${\bf P}^1$), 
which becomes useful here.  

We will be interested in a global version of these descriptions over
a complex {\em surface} 
$B$ so that our Calabi-Yau $Z$ can be described by a 
generalized Weierstrass equation in a ${\bf P}^2$ bundle $W$ over $B$
(note that in the case of the $B$-model the fibre ${\bf P}^2$ is actually
a weighted ${\bf P}_{2,1,1}$ with $y$ being a section of 
${\cal O}(2)$). 
In the following table we 
indicate the power $i$ of the line bundle ${\cal L}=K_B^{-1}$ over $B$ 
so that the given variables $x,y,z$ 
resp. coefficient functions $a,b,c,d,e$ are sections of ${\cal L}^i$.
{\footnotesize
\begin{center}
\begin{tabular}{c|c||c|c|c|c||c|c|c|c|c||c|c}
  & $Z$                                 & $z$ & $x$ & $y$ & $w$ &  $a$ & $b$& 
$c$& 
$d$ & $e$ & s &def$(K3)$ \\ \hline
A & $zy^2+x^3+axz^2+bz^3=0$             & $0$ & $2$ & $3$ & $ $ &  $4$ & $6$& $ 
$& 
$ $ & $ $ &  1 & 18 \\ \hline
B & $y^2+x^4+ax^2z^2+bx z^3+cz^4=0$     & $0$ & $1$ & $2$ & $ $ &  $2$ & $3$& 
$4$& 
$ $ & $ $ &  2 & 17 \\ \hline
C & $x^3+y^3+axyz+bx z^2 +cyz^2+dz^3=0$ & $0$ & $1$ & $1$ & $ $ &  $1$ & $2$& 
$2$& 
$3$ & $ $ &  3 & 16 \\ \hline 
D & ${\displaystyle{x^2+y^2+a zw +d z^2=0}, \atop 
     \displaystyle{w^2+xy+b xz+c yz+ez^2=0}}$          
                                        & $0$ & $1$ & $1$&  $1$ &  $1$ & $1$& 
$1$ 
& $2$ & $2$  & 4 & 15 \end{tabular}

\vskip 3 mm
{\bf Tab. 2} {\sl Generic fibrations with $A,B,C,D$ fibre types}
\end{center}}

Here we indicated also the number of cohomological inequivalent sections $s$. It 
is obtained by 
counting the solutions of the equation at the locus of the section $z=0$, taking 
into account 
the equivalence relations in the (weighted) projective spaces. E.g. for the 
$B$-fiber we get at 
$z=0$ the equation $y^2=x^4$, which has naively 8 solutions which lie however 
only 
in two equivalence 
classes in ${\bf P}(2,1,1)$: $[y,x]=[1,1]$ and $[-1,1]$ etc. . 
Note that always one of the $s$ 
sections is the 
zero section, while the others have rank 1 in the Mordell-Weil group, 
i.e. they 
generate 
infinitely many sections. For special points in the moduli space 
we can bring 
them 
to 
torsion points of finite order $s$, see below.        

Furthermore we indicated the number of complex structure deformations 
$d$ if we 
consider an elliptically fibered 
$K3$, with  ${\bf P}^1$ base. Then $a,b,c,\ldots $ are polynomials 
of order $2i$ in the ${\bf P}^1$ variables. 
The number of the parameters in the polynomials 
minus the $3$ parameters of the 
$SL(2,{\bf C})$ 
reparametrisations of the ${\bf P}^1$, minus one 
for the overall scaling of all variables 
gives the number of independent complex structure deformations. 
E.g. for the $B$-fibre $K_3$ we have $d=5+7+9-3-1=17$ etc. The  
Picard-group of the $K_3$ is generated by the $s$-sections plus the base. 
i.e. $\rho=s+1$ and in particular $\rho+d=20$. 

The construction of free ${\bf Z}_n$ actions on $Z$ will be closely
connected with shift symmetries, which are
free at least on the generic fibre. The existence of the shift 
symmetry implies that the
$n$ sections are at order $n$ points 
(${1\over n}$-periods), which are fixed under the monodromy; this 
in turn means specialization of the complex parameters\footnote{The spaces
then turn out to be singular; so this is only a rudimentary version 
of what we are actually going to do.}. The following
table shows for each of the models two consecutive specializations
and their ensuing monodromy enhancements [\ref{BKMT}].

{\footnotesize  
\begin{center}
\begin{tabular}{c||c|c||c|c}
          &        B                &          &        C                &         
 \\ \hline
specialization
&       $b=0$               & \& $c=h^4$ 
&       $b=c=0$             & \& $d=h^3$ \\ \hline
symmetry  & $ \matrix{(y,x,z)\cr
                     \downarrow \cr
                      (-y,-x,z)}$  &
               \&    $\matrix{(y,x,z)\cr
                     \downarrow \cr
                      (-y,zh,xh^{-1})}$ 
& $ \matrix{(y,x,z)\cr
                     \downarrow \cr
                      (y,\mu_3x,\mu_3^2z)}$  &
                \&    $\matrix{(y,x,z)\cr
                     \downarrow \cr
                      (x,zh,yh^{-1})}$ 
  \\ \hline
Mordell-Weil
& $Z_2$         &  $Z_2\times Z_2$
& $Z_3$         &  $Z_3\times Z_3$  \\ \hline
Monodromy
& $\Gamma_1(2)$         &  $\Gamma(2)$
& $\Gamma_1(3)$         &  $\Gamma(3)$  \\ \hline
def$(K3)$
& $10$         &  $6$
& $6$         &  $2$   \\ \hline
sing.
& $(A_1)^8$         &  $(A_1)^{12}$
& $(A_2)^6$         &  $(A_2)^8$   \\ \hline
\end{tabular}
\end{center}

\begin{center}
\begin{tabular}{c||c|c}
          &        D                &                     \\ \hline
specialization 
&       $b=c=d=0$             & \& $a=h^2$, $e=g^4$  \\ \hline
symmetry  & $ \matrix{(y,x,z,w)\cr
                     \downarrow \cr
                      (\mu_4^{-1}y,\mu_4 x, z,\mu_4^2 w)}$  &
               \&    $\matrix{(y,        x,   z,   w)\cr\downarrow \cr
                              (g^2h z,  hw,  g^{-1}x,g y)}$ 
  \\ \hline
Mordell-Weil
& $Z_4$         &  $Z_4\times Z_4$
\\ \hline
Monodromy
& $\Gamma_1(4)$         &  $\Gamma(4)$\\ \hline
 def$(K3)$
& $4$         &  $0$   \\ \hline 
sing
& $(A_1)^2\times (A_3)^4$         &  $ (A_1)^4\times (A_3)^4$   \\ \hline
\end{tabular}
\newline\nobreak
\vskip 3 mm
{\bf Tab. 3} {\sl Properties of specializations of the $B,C,D$-type 
fibres}
\end{center}}

The possible degenerations of the associated Weierstrass forms 
employing the Kodaira classification tables of singular fibres are 
also analysed [\ref{BKMT}]. I.e. the group is $\Gamma_1(k)$ $(\Gamma(k))$ 
for $k=9-j$ `the number' of the model related to $E_j$ 
(,i.e. $1,2,3,4$ for the $A,B,C,D$ model; 
cf. the Euler number later). The relation to the 
$E_j$ classification becomes apparent, if we blow up (down) a  
$(-1)$ curve in the base. In this case we create (or shrink) an
$E_j$ $j=8,\ldots 5$ del Pezzo surface [\ref{MVII}, \ref{KMV}], 
which can be geometrically described by $P_{3,2,1,1}(6)$, 
$P_{2,1,1,1}(4)$, $P^3(3)$ and $P^4(2,2)$. The local singularities
which have the same structure, were analysed in [\ref{reid}]. We assume that 
elliptic fibrations related to $E_j$ for $j<5$ 
with $9-j$ sections exist, but it is known that the  associated 
singularities mentioned above can not be described as complete 
intersections. The further specializations in the table, as
well as forms of the A-fibre, which have additional sections, are
singular and it is not known whether an extremal transition leads
to new fibre types.    

\subsection{Our procedure}

Our main goal will be to give a free ${\bf Z}_n$ action on $Z$.
One does not find free operations on the base $B$. Therefore the operation
over the fix-locus in the base must be free in the fibre. But this means that 
it is a shift by an $n$-torsion point. To force the existence of this shift
we have to use actually a fibration where such a shift exists globally (even
if only on a sublocus of moduli space).
So clearly the first
idea in choosing the $A,B,C,D$ models is to use the fibre shift
and to enhance it to an operation in the total Calabi-Yau by combining it
with an operation in the base. This will be necessary for two reasons: first
at the singular fibers the pure fibre operation ceases to be free, and secondly
to use just the fibre operation would us restrict to the first specialisation
locus given in the table which is singular\footnote{The unspecialised, generic
form of the $B,C,D$ fibrations are smooth for the bases $F_m$ for $m=0,1,2$ and
$dP_k$ for $k=0,1,2,9$ and then ten additional toric polyhedra described in
[\ref{koelman}].}. For example in the $B$ model
the involution $(y,x)\ra (-y,-x)$ in the coordinates 
(which for $b=0$ is identical to the shift provided by the second section, 
which is then an involution) 
does not necessarily restrict us to the 
$b=0$ locus {\em when combined} with a base operation. In other words
the involution $(y,x)\ra (-y,-x)$ does not exist on the fibre {\em per se},
but can exist, when combined with a base involution, on a subspace (not as
restrictive as the locus $b=0$; actually only some monomials {\em within} the 
coefficient functions $a,b,c$ are then forbidden) of the moduli space where the
generic member is still {\em smooth}. Note that even away from $b=0$ the 
coordinate involution maps still the two sections on another.

In general the full action, being given by these actions on the fibre
accompanied by suitable group actions on the basis, have, besides the 
requirement of {\em not restricting us to a non-smooth model 'above'} by the
condition of their definedness, 
also then to fulfill the further conditions to 
provide {\em free} group actions, which leave the {\em holomorphic three-form 
invariant}, as will be discussed further in section $4$.    

\subsection{Transitions}

Finally we discuss the relation between the $A,B,C,D$ smooth Weierstrass 
Calabi-Yau three folds. 
If we force a second section for the $A$-type Weierstrass fibration 
$Z$ 
it takes the general  form $y^2=4(x+f)(x^2-f x-g)$, where $f,g$ are sections of 
${\cal L}^2,{\cal L}^4$. From 
$$g_2=16 \cdot 2^{2/3}(f^2+g),\quad g_3=64 fg, \quad  \Delta=(g-2 f^2)^2 (f^2+4 
g)$$ 
wee see using Kodairas classifications of singular fibers that it acquires an 
$A_1$ fiber over a divisor $B'$ in the class $[B'_{A_1}]=-4 K_B=4 c_1$ 
and the calculational framework [\ref{FMW}] needs modifications. 
The smooth A-fibre Weierstrass model had Euler number $\int_Z c_3(Z)=-60 
\int_B c_1^2(B)$ and if we resolve the singularities over the divisor $B'$ we 
get 
an 
extremal transition\footnote{Take type IIA on the same 
Calabi-Yau space, then we get physical interpretation of the transition. 
Unhiggsing 
of a $SU(2)$ with $g(B')$ hypermultiplets in the adjoint 
[\ref{KM},\ref{KPM}].} 
{}From $Z$ 
to a model $\hat Z$, whose change in the Euler number 
[\ref{KLRY}] is given by 
$\delta={\rm cox}(G) {\rm rank}(G) \int_{B'} c_1(B')$. 
By the adjunction formula 
we calculate 
$\delta=2 \int_{B'} c_1(B')=24 \int_B c_1^2(B)$, 
which yields $\int_{\hat Z} 
c_3(\hat Z)=
-36\int_B c_1^2(B)$. This model is up to birational transformations 
the elliptic fibration with $B$-fibre type over the same base, 
whose Euler number depends on the base cohomology in 
precisely this way [\ref{KLRY}]. 
Similarly forcing three sections produces an $A_2$ 
singularity over $[B']=-3 K_B$ [\ref{BKMT}], 
and the extremal transition leads with 
$\delta=6 \int_{B'} c_1(B')=
36 \int_B c_1^2(B)$ to $\int_{\hat Z} c_3(\hat Z)=-24\int_B c_1^2(B)$, 
the $C$-fibre type. Finally forcing four sections leads by a slightly more 
complicated transition
to the $D$-fibration.

\subsection{The cohomology of the spaces}  
\label{cohomology}
The different fibration structures leads in the following way to the 
cohomological data of $Z$ (we give, as an example, 
a consideration of the $B$-model, cf. for the $A$-model [\ref{FMW}];
unspecified Chern classes will always refer to the base $B$). 
 
As noted $z,x,y$ can be thought of as homogeneous coordinates on $W$ 
respectively globally as sections of line bundles 
${\cal O}(1), {\cal O}(1)\otimes {\cal L}$ and 
${\cal O}(1)^2\otimes {\cal L}^2$
whose first Chern classes are given by $r$, $r+c_1$, $2r+2c_1$
with $c_1({\cal O}(1))=r$. 
The cohomology ring of $W$ is generated
by $r$ with the relation $r(r+c_1)(2r+2c_1)=0$ expressing the fact that 
$z,x,y$ have no common zeros. Since the $B$-model is defined by the 
vanishing 
of a section of ${\cal O}(1)^4\otimes {\cal L}^4$, which is a 
line bundle over 
$W$ with first Chern class $4(r+c_1)$,
the restriction from $W$ to $Z$ is effected by multiplying by this 
Chern class,
so that $c(W)=(1+4r+4c_1)c(Z)$. One can then simplify
$r(r+c_1)(2r+2c_1)=0$ to $r(r+c_1)=0$ in the cohomology ring of $Z$ 
and finds
for the total Chern class of $Z$ (B-model)
\beqa
c(Z)=c(B)\frac{(1+r)(1+r+c_1)(1+2r+2c_1)}{1+4r+4c_1}
\eeqa
Using $r^2=-rc_1$ and taking into account that the divisor $(z=0)$ of
the section $z$ of the line bundle ${\cal O}(1)$ of class $r$
shows that $r=\sigma_1+\sigma_2$ (as $z=0$ implies $y^2=x^4$, leading to
$(x,y)=(i,1)$ and $(i,-1)$) we find\footnote{for the Euler number we give the 
cohomology class which has to be integrated over the base}
\beqa
c_2(Z)&=&c_2+6(\sigma_1+\sigma_2) c_1 +5c_1^2\nonumber\\
c_3(Z)&=&-36 c_1^2
\eeqa
The computations for the $C$ and $D$ model can be done the same way and 
one finds for the cohomological data of $Z$ (if one defines the
sum of the sections $\Sigma=\sum_{i=1}^k \sigma_i$, so that one
has $r=\Sigma$ for the $B,C,D$ models in contrast to the $r=3\sigma$
of the $A$ model (note that the 'simplification relation' is $r(r+c_1)=0$ 
in the $B,C,D$ models
in contrast to the $r(r+3c_1)=0$ in the $A$ model, the only case
where a section occurs with a multiplicity (namely $3$) in $r$); so in
all cases the adjunction relations $\sigma_i^2=-\sigma_i c_1$ are 
satisfied, as they should)

\begin{center}
\begin{tabular}{c|c|c|c}
  & $G$   & $c_3(Z)$   & $c_2(Z)$               \\ \hline
A & $E_8$ & $-60c_1^2$ & $c_2+12\sigma c_1 +11c_1^2$ \\ \hline
B & $E_7$ & $-36c_1^2$ & $c_2+ 6\Sigma c_1 + 5c_1^2$  \\ \hline
C & $E_6$ & $-24c_1^2$ & $c_2+ 4\Sigma c_1 + 3c_1^2$  \\ \hline
D & $D_5$ & $-16c_1^2$ & $c_2+ 3\Sigma c_1 + 2c_1^2$ 
\end{tabular}
\vskip 2 mm
{\bf Tab. 4} {\footnotesize \sl Chern classes of the $A,B,C,D$ models }
\end{center}

What concerns the volume form, e.g. for the 
$B$-fibre we have $-18 \int_Zr c_1^2=-18 \int_Zc_1^2 \Sigma=-36 \int_Bc_1^2$.  

So one sees that in general (with $k$ the 
'number' of the model (cf. the monodromy groups above) and 
$h$ is the Coxeter number of $G$)
\beqa
c_2(Z)&=&c_2+\frac{12}{k}\Sigma c_1 + (\frac{12}{k}-1)c_1^2\nonumber\\
c_3(Z)&=&-2h c_1^2 
\eeqa

\subsection{Examples}

\subsubsection{Over Hirzebruch bases}
\label{f0examples}
Let us consider as an example the $B$-type fibration over $B=F_0$. 
As on reads off from the table of weight given above, $a,b,c$ are sections 
of line bundles corresponding to polynomials of bi-degree $(4,4)$, $(6,6)$ and 
$(8,8)$ 
over $F_0={\bf P}^1\times {\bf P}^1$, which yields 
$h_{2,1}=5^2+7^2+9^2-3-3-1=148$, 
where the subtraction comes from the two ${\rm SL}(2,{\bf C})$ automorphism on 
the 
two ${\bf P}^1$ 
and the rescaling. Of the $(1,1)$-forms two come from base and 
two from the two sections. Over $F_2$ our Calabi-Yau is the 
${\bf P}_{1,1,2,4,8}(16)$ just as in the $A$ fibre case one had the
${\bf P}_{1,1,2,8,12}(24)$.

\subsubsection{Over rational elliptic bases}

Let us consider as an example, which becomes important later in section (4.2), 
the case $B=dP_9$. 
Let us first recall that we denote by $dP_9$ the rational elliptic surface
being given by the projective plane $P^2_{x,y,z}$ 
blown up in the nine intersection points of two cubics. The later will be 
denoted 
by
$C,C^{\prime}$. This representation $sC(x,y,z)+tC^{\prime}(x,y,z)=0$ 
(with $s,t$ coordinates on $P^1$) of $dP_9=
{\tiny \left[\begin{array}{c|c}P^2&3\\P^1&1\end{array}\right]}$ shows
at the same time the elliptic fibration, with $C$-type elliptic fibre. 
This may also be called  $\frac{1}{2}K3$ for the $K3=
{\tiny \left[\begin{array}{c|c}P^2&3\\P^1&2\end{array}\right]}$; it has
$h^{2,0}=0$ and $h^{1,1}=10$ and $8$ complex deformations. These can be 
counted either as $8$ points (the ninth is determined as they have to 
add up to zero in the group law) on $P^2$, leading to $8 \times 2 - 8=8$ 
parameters, where
the each point contributes $2$ degrees of freedom and the rescaling comes from 
${\rm PGl}(3,{\bf C})$. Alternatively one can count the monomials as $10\cdot 2 
-8-3-1=8$. 
{}From this surface one can build the Calabi-Yau space
${\tiny 
\left[\begin{array}{c|cc}P^2&3&0\\P^1&1&1\\P^2&0&3\end{array}\right]}$.
Note that our earlier constructed Calabi-Yau spaces
have such a fibre product structure in case the base is a $dP_9$: the fact
that the coefficient functions $f$ and $g$ are sections of a line bundle
${\cal L}$ of class $c_1(dP_9)=f$ 
(here $f$ denotes the elliptic fibre of $dP_9$) 
means that they are linear in the base $P^1$ and 
{\em independent of the fibre direction} (so having polar divisor $f$).
Note further that these Calabi-Yau spaces $Z$ when restricted to the base
$P^1$ (of the base (i.e. horizontal) $dP_9$) 
have again the structure of a (vertical) $dP_9$, with correspondingly changed
elliptic fibre type.  This is
the well known representation $dP_9=\frac{1}{2}K3$: if $B$ would be 
one-dimensional one would get a $K3$ for $c_1({\cal L})=c_1(P^1)=2$ which
would double the degrees of the coefficient functions. So changing
{}From $K3$ to $\frac{1}{2}K3$ means in the $B$ model to consider
$y^2+x^4+a_2(s,t)x^2z^2+b_3(s,t)xz^3+c_4(s,t)z^4=0$.

Note that although the $K3$ in the $A,B,C$ model have because of
the added sections a decreasing number of deformations, 
one has nevertheless by going to the rational elliptic surfaces, i.e.
the corresponding $\frac{1}{2}K3$, always the $8$ deformations as one
easily convinces oneself in dividing the mentioned degrees by two (for
example $3+4+5-3-1=8$ in the $B$ model); i.e.
one always has for them the same characteristics $h^{1,1}=10=\rho$ 
and $8$ complex deformations, the difference between the $A,B,C$ models
and further the mentioned $C'$ model 
${\tiny \left[\begin{array}{c|c}P^2&3\\P^1&1\end{array}\right]}$,
whose ambient space is in contrast to our $C$ model a {\em constant} ${\bf 
P}^2$-fibration
over the base, being given by the type of the elliptic fibre in which different
group actions are manifest.

Note that the Euler number vanishes 
for a $dP_9$ base in our examples anyway (because of $c_1^2=f^2=0$) as
it should for double elliptic fibrations (the two sets of $12$ points
in the base $P^1$ where both the elliptic fibres of the two $dP_9$ 
fibre factors degenerate are generically disjoint).

\section{The bundles}

\subsection{The spectral cover description}

Let us recall the
idea of the spectral cover description of an $SU(n)$ bundle $V$: 
one considers the bundle
first over an elliptic fibre and then pastes together these descriptions
using global data in the base $B$. Now over an elliptic fibre $E$ 
an $SU(n)$ bundle $V$ over $Z$ (assumed to be fibrewise semistable)
decomposes as a direct sum of line bundles of degree zero; this is 
described as a set of $n$ points which sums to zero. If you now let 
this vary over the base $B$ this will give you a hypersurface 
$C \subset Z$
which is a ramified $n$-fold cover of $B$. If one denotes the cohomology
class in $Z$ of the base surface $B$ 
(embedded by the zero-section $\sigma_1$) by 
$\sigma \in H^2(Z)$ one finds that the globalization datum
suitable to encode the information about $V$ is given by a class 
$\eta \in H^{1,1}(B)$ with\footnote{$C$ is given as a locus $w=0$ with $w$
a section of ${\cal O}(\sigma)^n\otimes {\cal M}$ 
with a line bundle ${\cal M}$ of class $\eta$.}
\beqa
C=n\sigma + \eta
\eeqa

The idea is then to trade in the $SU(n)$ bundle $V$ over $Z$,
which is in a sense essentially a datum over $B$, 
for a line bundle $L$ over
the $n$-fold (ramified) cover $C$ of $B$: one has 
\beqa
V=p_*(p_C^*L\otimes {\cal P})
\eeqa
with $p:Z\times_B C\ra Z$ and $p_C: Z\times_B C\ra C$ the projections
and ${\cal P}$ the
global version of the Poincare line bundle over $E\times E$ 
(actually one uses a symmetrized version of this), i.e. 
the universal bundle
which realizes the second $E$ in the product as the 
moduli space of degree zero
line bundles over the first factor.

A second parameter in the description of $V$ is given by a
half-integral number $\lambda$ which occurs because one gets 
{}From the condition $c_1(V)=\pi_*(c_1(L)+\frac{c_1(C)-c_1}{2})=0$ 
that with
$\gamma \in ker\, \pi_*:H^{1,1}(C)\ra H^{1,1}(B)$ 
one has
\beqa
c_1(L)=-\frac{1}{2}(c_1(C)-\pi_{*}c_1)+\gamma
\eeqa
where $\gamma$ is being given by ($\lambda \in \frac{1}{2}{\bf Z}$)
\beqa
\gamma=\lambda(n\sigma-\eta+nc_1)
\eeqa
as $n\sigma_1|_C-\eta+nc_1$ is the only generally given class which
projects to zero. As we are in a different set-up than
[\ref{FMW}] one has actually a further possibility which is
described in section 5.

\subsection{The generation number}

Considering the chiral matter content one finds then 
as the number of net generations [\ref{C}]
\beqa
\frac{1}{2}c_3(V)=\lambda \eta (\eta-nc_1)
\eeqa
Note that because of the decomposition
\beqa
{\bf 248}&=&({\bf 3},{\bf 27})\oplus({\bf \bar{3}},{\bf \bar{27}})\oplus
({\bf 1},{\bf 78})\oplus ({\bf 8},{\bf 1})\nonumber \\
 &=& ({\bf 4},{\bf 16})\oplus ({\bf \bar{4}},{\bf \bar{16}})\oplus
({\bf 6},{\bf 10})\oplus ({\bf 1},{\bf 45})
\oplus ({\bf 15},{\bf 1})\nonumber \\
 &=& ({\bf 5},{\bf 10})\oplus ({\bf 10},{\bf \bar{5}})\oplus
({\bf \bar{5}},{\bf \bar{10}})\oplus ({\bf \bar{10}},{\bf 5})\oplus
({\bf 1},{\bf 24})\oplus ({\bf 24}, {\bf 1})
\eeqa
one has, in order to get all the relevant fermions, 
in the case of an $SU(5)$ bundle - unlike the case of an $SU(3)$
or $SU(4)$ bundle - to consider also
the $\Lambda ^2 V={\bf 10}$ to get the ${\bf \bar{5}}$ part of the 
fermions ${\bf 10} \oplus {\bf \bar{5}}$; but the ${\bf 10}$ and the 
${\bf \bar{5}}$ will come in the same number of families by anomaly 
considerations.

\subsection{The cohomology of the bundles}

Note that in order to apply the formalism of [\ref{FMW}], to compute
the generation number of $V$ and to make 
the checks on the effectiveness of the five-brane class below
we have to take into account that we are now working in Calabi-Yau
spaces with a different representation of the elliptic fibre $E$. 
This changes the cohomological data of $Z$ itself 
(as described above) 
but could also change the expression for $c_2(V)$ and $c_3(V)$ 
if you look at
Grothendieck-Riemann-Roch 
\beqa
\pi_*(e^{c_1(L)}Td(C))=ch(V)Td(B)
\eeqa
which shows an influence in $c_2(V)$
of the change of $c_2(Z)$ relative to the $A$ model:
as $c_2(Z)$ occurs in the Chern classes $c_1(C)=-(n\sigma_1+\eta)$ 
and $c_2(C)=c_2(Z)+c_1^2(C)$ of the spectral cover (note that the
whole spectral cover construction always uses only the zero section
$\sigma_1$) one gets the following. The new term to consider
is the push-down (from $C$ to $B$) 
of $\Sigma$ (coming from $c_2(Z)$ in $c_2(C)$; the complete term
there is a $\frac{12}{k}\Sigma c_1$ inside $c_2(Z)$, times $\frac{1}{12}$
because of $Td(C)$); 
this can also be understood as the push-down from $Z$ to $B$ of
$C\Sigma$ which is $(n\sigma_1+\eta)\sum_{i=1}^k \sigma_i=
(-nc_1+\eta)\sigma_1+\eta\sum_{i=2}^k \sigma_i=
-nc_1\sigma_1+\eta\sum_{i=1}^k \sigma_i$; this is pushed down to
$-nc_1+k\eta$, so the whole term inside $c_2(Z)$ goes to
$\frac{1}{12}(-n\frac{12}{k}c_1+12\eta)c_1$. 
The second summand here is especially
important as it gives\footnote{By doing the GRR for $\pi:C\ra B$
we actually compute only $c_2(V)|_{\sigma_1}$; so the term '$-\eta c_1$' 
in the expression for the restriction comes because of 
$\sigma_1|_{\sigma_1}= -c_1|_{\sigma_1}$ (from adjunction) from
a term $\eta \sigma_1$ in the total expression for $c_2(V)$. 
This, and the further statement that in the cases with 
changed elliptic fibre the element which restricts to $\sigma_1$ does not
involve the other $\sigma_i$ (say in the combination $\Sigma$), 
can be checked from the corresponding GRR formula for 
$p:Z\times _B C\ra Z$ (this is described in [\ref{C}]); there one sees 
that the corresponding term comes from the Poincare bundle, which involves
only $\sigma_1$.}
the first term (which, as we see, remains unchanged
compared to the $A$ model as $k$ cancels out)
in the final result for 
$c_2(V)=\eta\sigma_1+\omega$ 
(where $\omega\in H^4(B)$, pulled back to $Z$.

Now let us give the second Chern class for $V$ 
(one has $\pi_*(\gamma^2)=-\lambda^2 n (\eta(\eta-nc_1)$):
\beqa
c_2(V)&=&\frac{\frac{12}{k}k}{12}\eta \sigma_1 -
\frac{n^3-(2(a-b)-1)n}{24}c_1^2
-\frac{n}{8}\eta (\eta-nc_1)-\frac{1}{2}\pi_*(\gamma^2)) \nonumber\\
 &=&\eta \sigma_1 -\frac{n^3-n}{24}c_1^2+
\frac{1}{2}(\lambda^2-\frac{1}{4})n\eta (\eta-nc_1) 
\eeqa
i.e. there is no correction 
as $a-b=1$ in $c_2(Z)=c_2+a\Sigma c_1 +bc_1^2$.
Note that $c_3(V)$ is not changed too.

\subsection{The parabolic approach}

Of course this is no accident that the bundle does not see the
changed fibre. One has a second approach to describe bundles
and compute their Chern classes [\ref{FMW}],[\ref{A}]. In the parabolic
approach one starts with an unstable bundle on 
a single elliptic curve $E$.
For this one fixes a point $p$ on $E$ with the associated 
rank $1$ line bundle ${\cal O}(p)=W_1$. 
Rank $k$ line bundles $W_k$ are then
inductively constructed via the unique non-split extension 
$0\rightarrow 
{\cal O}\rightarrow W_{k+1}\rightarrow W_k\rightarrow 0$. If 
one writes the dual of $W_k$ as $W_k^*$ 
then the unique (up to translations 
on $E$) minimal unstable bundle with 
trivial determinant on $E$ is given by 
$V=W_k\oplus W^*_{n-k}$. 
This can be deformed by an element of
$H^1(E,W^*_k\otimes W^*_{n-k})$ to a stable bundle 
$V'$ which fits then into 
the exact sequence $0\rightarrow W^*_{n-k}\rightarrow V'\rightarrow 
W_k\rightarrow 0$.

To get a global version of this construction on replaces the $W_k$ 
by their global versions, i.e. 
replace ${\cal O}(p)$ by ${\cal O}(\sigma_1)$.
The global versions of $W_k$ are inductively constructed 
by an exact sequence 
$0\rightarrow 
{\cal L}^{n-1}\rightarrow W_k\rightarrow W_{k-1}\rightarrow 0$. 
Using the fact that one can 
globally twist by additional data coming from the base, 
i.e. line bundles
${\cal M}$ and ${\cal M}'$ on $B$, one finds the unstable $SU(n)$ bundle 
\beqa
V=W_k\otimes {\cal M}\oplus W_{n-k}^*\otimes {\cal M}'
\eeqa
with $W_k=\bigoplus_{a=0}^{k-1}{\cal L}^a$, $W_{n-k}^*=
\bigoplus_{b=0}^{n-k-1}{\cal L}^{-b}$ and ${\cal M},{\cal M}'$ are 
constrained so that $V$ has trivial determinant. Note that the 
specification of the twisting line bundle ${\cal M}$ corresponds to the
specification of $\eta$ in the spectral cover approach (cf.[\ref{FMW}],
[\ref{A}]). Since the topology of $V$ is invariant under deformations one 
computes Chern classes simply of $V$ using 
\beqa
c(V)=\prod_{a=0}^{k-1}(1+c_1({\cal L}^a)+c_1({\cal M}))
\prod_{b=0}^{n-k-1}(1+c_1({\cal L}^{-b})+c_1({\cal M}'))
\eeqa   
which is independent of the fibre type ($A,B,C,D$). 

\subsection{Bundle moduli and index theorems}

Let us also point out that the ${\bf Z}_2$ equivariant index $I=n_e-n_o$ 
of [\ref{FMW}], 
counting the bundle moduli which are even respectively odd 
under the $\tau$-involution,  
does not change under exchanging the elliptic fibre of $Z$. Recall 
that $I$ can be interpreted as giving essentially the holomorphic Euler 
characteristic of the spectral surface [\ref{CD}] which is 
\beqa
1+h^{2,0}(C)-h^{1,0}(C)&=
&\frac{c_2(C)+c_1^2(C)}{12}|_C=\frac{c_2(Z)C+2C^3}{12}\nonumber\\
&=&n+\frac{n^3-n}{6}c_1^2+\frac{n}{2}\eta(\eta-nc_1)+
\frac{\frac{12}{k}k}{12}\eta c_1
\eeqa
Now identifying the number of local complex 
deformations $h^{2,0}(C)$ of $C$ with $n_e$
respectively the dimension $h^{1,0}(C)$ of $Jac(C):= Pic_0(C)$
with $n_o$, one finds
\beqa
I=n-1+\frac{n^3-n}{6}c_1^2+\frac{n}{2}\eta(\eta-nc_1)+\eta c_1.
\eeqa

\subsection{Restrictions on the bundle parameters}

After these introductory remarks we proceed to the actual 
construction of the examples.
This will be done in two steps: first we specify the vector bundle 
('above') over $Z$ and then we make explicit the involution. 
For this let us recapitulate what are the requirements: we are searching
an $SU(5)$ resp. $SO(10)$ bundle $V$ with $6$ resp. $12$ 
net-generations on a Calabi-Yau
admitting a freely acting ${\bf Z}_2$ which respects 
the holomorphic three-form. The conditions concerning the Calabi-Yau $Z$
will be treated in the next section. 
The conditions concerning the bundle $V$ amount to the following.

\subsubsection{Restriction on $\lambda$}

{}From
\beqa
c_1(L)=n(\frac{1}{2}+\lambda)\sigma +(\frac{1}{2}-\lambda)\eta+
(\frac{1}{2}+n\lambda)c_1
\eeqa
one sees that the easiest (and the only general) way to fulfill the 
requirement of integrality\footnote{Note that these conditions assure the 
integrality
of $c_2(V)$ too.} for $c_1(L)$ is to require $\lambda$
to be strictly half-integral for $n$ odd resp. to be strictly
integral and $\eta\equiv c_1 \, {\rm mod} \, 2$ for $n$ even.
More exotic possibilities involve $\lambda=\frac{1}{2n}$ for $n$ odd
and $\eta\equiv 0\, {\rm mod}\, n$ [\ref{A}] or for example 
$\lambda=\frac{1}{4}$ for $n=4$ and $\eta=2c_1\, {\rm mod}\, 4$.

Combined with the searched for generation number obvious restrictions 
result on the possible $\lambda$ (for example $\lambda=\pm 1/2,\, \pm 3/2$ 
where one has then to construct on the Calabi-Yau 'above' (before the modding)
a model with $\eta(\eta-5c_1)=\pm 12$ or $\pm 4$ in the $SU(5)$ case).

\subsubsection{The upper bound on $\eta$}

The essential restrictions on $V$ come from bounds on the $\eta$ class. 
The upper bound comes from the anomaly cancellation condition
$c_2(Z)=c_2(V_1)+c_2(V_2)+W$ (we have here $V_2=0$)
giving the effectiveness restriction $c_2(V)\le c_2(Z)$ on the
five-brane class $W=W_B+a_f \, F$; clearly here a second effectiveness
condition is emerging for $a_f$: it has to be (integral and) 
non-negative\footnote{in the case of 
a Hirzebruch surface $F_k$ with $k \ge 3$ it is sufficient to have
non-negativity of $a_f-W_Bc_1$ (cf. [\ref{DLOW}])}. 
Now remember that in the final result for 
$c_2(V)=\eta\sigma_1+\omega$ 
(where $\omega\in H^4(B)$, pulled back to $Z$) the number' $k$ 
of the model ($1,2,3,4$ for $A,B,C,D$) cancelled out 
leaving the $A$ model result unchanged. On the other hand note that we
have a corresponding decomposition 
$c_2(Z)=\frac{12}{k}\Sigma c_1 + (c_2+(\frac{12}{k}-1)c_1^2)$).
This is the term responsible
for the upper bound $\eta\sigma_1\le \frac{12}{k}\Sigma c_1$, i.e.
$\eta\le \frac{12}{k}c_1$ (which is thus much sharper than in the
$A$ model) from the effectiveness restriction on the five-brane class.

\subsubsection{The lower bound on $\eta$}

This is a bound on 'how much instanton number has to be turned on
to generate/fill out a certain $SU(n)$ bundle', or speaking
in terms of the unbroken gauge group $G$ (the commutator of $SU(n)$ in
$E_8$) a condition 'to have no greater unbroken gauge group than a certain $G$'.

Let us recall the situation in six dimensions. There the easiest duality
set-up is given by the duality of the heterotic string on $K3$ with
instanton numbers $(12-m,12+m)$ (and no five-branes) with $F$-theory on
the Hirzebruch surface $F_m$ [\ref{MVII}]. The gauge group there is 
described by the singularities of the fibration and a perturbative
heterotic gauge group corresponds to a certain degeneration over the 
zero-section $C_0$ (of self-intersection $-m$): for example to get an 
$SU(3)$ one needs a certain $A_2$ degeneration over $C_0$ available 
first for $m=3$; in general this means that the discriminant divisor 
$\Delta=12c_1(F_m)$ has a component $\delta(G)C_0$ where $\delta(G)$ is
the vanishing order of the discriminant (equivalently the Euler number of
the affine resolution tree of the singularity), giving also the relation
$m\le \frac{24}{12-\delta(G)}$ for the realization over a $F_m$ to have no
singularity worse than $G$. The last relation follows (cf. [\ref{R}])
from the fact that after taking the $C_0$ component with its full 
multiplicity $\delta(G)$ out of $\Delta$ the resulting 
$\Delta^{\prime}=\Delta-\delta(G)C_0$ has transversal intersection 
with $C_0$ and so $\Delta^{\prime}\cdot C_0\ge 0$, leading with $c_1(F_m)
=2C_0+(2+m)f$ to the mentioned result.

So the instanton number $12-m$ to give a $G$ gauge group has to be
$12-m\ge 12-\frac{24}{12-\delta(G)}=(6-\frac{12}{12-\delta(G)})c_1(B_1)$ 
with $B_1$ the common $P^1$ base of the heterotic $K3$ resp. the $F_m$.  From 
this it was conjectured in [\ref{R}] (see also [\ref{BM}]) that a similar
effectiveness bound could
in four dimensions look like the generalizations of both sides of the 
six-dimensional bound, i.e. in view of the fact that the $(12-m,12+m)$
structure generalizes in four dimensions to $\eta_1=6c_1-t, \eta_2=6c_1+t$
(for this cf. the anomaly cancellation condition $c_2(V_1)+c_2(V_2)+a_fF=
c_2(Z)$ and its component $\eta_1\sigma+\eta_2\sigma=12c_1\sigma$ 
concerning the classes not pull-backed from $H^4(B)$ for the case of
an $A$ model with $W_B=0$)
\beqa
\eta_1\ge (6-\frac{12}{12-\delta(G)})c_1
\eeqa

So finally one has to fulfill the lower bound 
$\eta \ge \frac{30}{7} c_1$ resp. $\eta \ge \frac{18}{5} c_1$ 
in the case of an $SU(5)$ resp. $SO(10)$ unbroken 
gauge group (the coefficients
come from the expression $6-\frac{12}{12-\delta}$ in [\ref{R}]
with $\delta$ the vanishing order $5$ resp. $7$ of the discriminant).
As our breaking scheme involves one resp. two ${\bf Z}_2$'s (and so
the $B$ resp. $D$ model) for unbroken $SU(5)$ resp. $SO(10)$ we see that,
in view of the upper bound $\eta\le 3c_1$ in the $D$ model case,
we are left with the possibility of breaking $SU(5)$ via the $B$ model.

\subsection{Restricting attention to $V_1$}

When giving in the following subsection our examples we will restrict
attention to the visible sector, i.e. to the '$1$' sector in the 
bundle embedding $(V_1,V_2)\in E_8\times E_8$, and will neglect the hidden
'$2$' sector concerning $V_2$. This is justified up to the occurrence 
of the compensating five-brane class in
\beqa
c_2(V_1)+c_2(V_2)+W=c_2(Z)
\eeqa
which connects the two sectors. As we will consider a $B$ type model 
of $Z$ we will have $c_2(Z)=6c_1\sigma_1+6c_1\sigma_2$. Now 
the decomposition (suitable pull-backs understood)
$(\oplus_i \sigma_i H^2(B))\oplus H^4(B)$
leads for the parts not coming from $H^4(B)$ to
\beqa
\eta_1\sigma_1+\eta_2\sigma_j+W_B=6c_1\sigma_1+6c_1\sigma_2
\eeqa
where $j=1$ or $2$ according to whether we choose to build $V_2$ using
the same section $\sigma_1$ as for $V_1$ or the other one, i.e. $\sigma_2$
(the spectral cover construction uses in its essential parts just the
possible globalization of a point $p$ on an elliptic fiber curve, relative
to which degree zero line bundles are transmuted into divisors $Q-p$ for
another point $Q$; there is no geometric distinction which forces one of
the sections to be called the zero-section with respect to the group law).

So we have actually two possible strategies for saturating on the LHS of
the last equation the $6c_1\sigma_2$ of the RHS: either we put it in 
a "$W_B^{(2)}$" or we choose $j=2$ and $\eta_2=6c_1$. So let us now
also introduce the decomposition $W_B=W_B^{(1)}+W_B^{(2)}$ by which we 
mean $W_B^{(2)}=6c_1\sigma_2$ for $j=1$ (actually we will then not even 
turn on at all a second bundle, i.e. $\eta_2=0,\lambda_2=0$, as the lower
bound leaves no space for this) 
respectively $W_B^{(2)}=0$ for $j=2$ in
which latter case also $\eta_2=6c_1$ is understood. We do so
just for the sake of simplicity; of course one can easily 
elaborate a more general scheme for $V_2$
lying between the extreme cases $j=1$ (i.e. no bundle constructed over
$\sigma_2$ and further even $\eta_2=0, \lambda_2=0$) 
together with a maximal $W_B^{(2)}=6c_1\sigma_2$ on the one hand or 
$j=2$ and a maximal $\eta_2=6c_1$ and no $W_B^{(2)}$. This would repeat 
just the treatment of the "$1$" sector apart from the fact that now both
these sectors are not disentangled because there is still the $H^4(B)$
part of the anomaly equation giving the number of five-branes wrapping
the elliptic fibre
\beqa
-\frac{n_1^3-n_1}{24}c_1^2+
\frac{1}{2}(\lambda_1^2-\frac{1}{4})n_1\eta_1(\eta_1-n_1c_1)+
\;\;\;\;\;& &\nonumber\\
-\frac{n_2^3-n_2}{24}c_1^2+
\frac{1}{2}(\lambda_2^2-\frac{1}{4})n_2\eta_2(\eta_2-n_2c_1)+a_fF
&=&c_2+5c_1^2
\eeqa
Decompose now $a_f=a_f^{(1)}+a_f^{(2)}$. Our procedure below will be to
solve for $a_f^{(1)}$ (which then has to be 
integral and non-negative\footnote{of course one could imagine a more general
procedure which demands these conditions only for the total $a_f$ as only this
has an absolute meaning}) in
\beqa
-\frac{n_1^3-n_1}{24}c_1^2+
\frac{1}{2}(\lambda_1^2-\frac{1}{4})n_1\eta_1(\eta_1-n_1c_1)+a_f^{(1)}F
=c_2+5c_1^2
\eeqa
thereby saturating the base part of $c_2(Z)$. The other $a_f^{(2)}$ 
(in the case where a second bundle is turned on)
\beqa
a_f^{(2)}=\frac{n_2^3-n_2}{24}c_1^2-
\frac{1}{2}(\lambda_2^2-\frac{1}{4})n_2\eta_2(\eta_2-n_2c_1)
\eeqa
will take care of itself: the non-negativity and integrality 
of $a_f^{(2)}$ are easily accomplished by choosing $n_2=3$ or $5$
and $\eta_2=6c_1, \lambda_2=1/2$.

So the upshot is we have now reduced the discussion to the '$(1)$' sector
and shall omit below the $(1)$ index, it being understood that a 
$W_B^{(2)}=6c_1\sigma_2$ (and $\eta_2=0,\lambda_2=0$)
or an $\eta_2=6c_1$ runs in the second sector.

\subsection{Examples}

So one has to find an $\eta$
in the strip 
$\frac{30}{7}c_1\le \eta \le 6c_1$ with $\eta(\eta-5c_1)=\pm 12$
(for $\lambda=\pm \frac{1}{2}$) resp. $\eta(\eta-5c_1)=\pm 4$ 
(for $\lambda=\pm \frac{3}{2}$) with $a_f$ non-negative (we will be
considering only $F_0,\, F_1,\, F_2$ and the $dP_k$'s) where
$a_f=c_2+10c_1^2$ for $\lambda=\pm \frac{1}{2}$ resp.
$a_f=c_2+10c_1^2\mp 20$ for $\lambda=\pm \frac{3}{2}$ (so this
only excludes for $\lambda=+\frac{3}{2}$ the $dP_9$; of course
one can work in that case simply still with $\lambda=-\frac{3}{2}$
getting $-6$ net-generations).

Now let us specify the $SU(5)$ bundle over $Z$.
Recall that the Hirzebruch surfaces $F_m$ are $P^1_{(2)}$ fibrations
over $P^1_{(1)}$ possessing a section of self-intersection $-m$ and have
$c_1(F_m)=(2,2+m)$ in a basis (for the effective cone) $(b,f)$ with
$b^2=-n$ and $f$ the fiber. The data which specify our bundle over $F_1$
are now given by
\beqa
\lambda=+\frac{3}{2}, \; \eta= (11,15) \nonumber
\eeqa

Note that besides leading to a net-number of $6$ generations
these data fulfill the  series of further conditions mentioned above
with $W=(1,3)+64 \, F$. This is (in the given set-up) the 
only solution on a Hirzebruch surface $F_r$, $r=0,1,2$. 
Over the del Pezzo surfaces
$dP_k$ there are many more possibilities (here of course 
the example over $F_1$
reoccurs; we are using the basis $(l,(E_i)_{i=1...k})$ where for example
$c_1=(3,-1^k)$; here one has to keep in mind that the classes $l-E_i$ and 
$l-E_i-E_j$ when
$n\ge 2$, $2l-\sum_{j=1}^5E_{i_j}$ when $n\ge 5$, 
$3l-2E_i-\sum_{j=1}^5E_{i_j}$ when $n\ge 7$ and 
$4l-2\sum_{j=1}^3E_{i_j}-\sum_{m=1}^5E_{i_m}$ when $n\ge 8$
are effective when verifying bounds on
$\eta$). Note that the five-brane class $W=W_B+a_f\, F$ is given by
$W_B=6c_1-\eta$ and $a_f=93-9k$ for $\lambda=\pm \frac{1}{2}$ resp.
$a_f=93-9k \pm 20$ for $\lambda=\pm \frac{3}{2}$. Besides a long list
of $6$ generation models which can be given for the del Pezzo surfaces up to 
$dP_8$
(which we will not list as we will have no suitable involution on $Z$),
on $dP_9$ ($C'$-type; we will give in the next section a fixpoint free 
involution for $dP_9^B$ of $B$-type), one sees that one has no 
relevant examples, as the bounds ${30\over 7} c_1\le 
\eta\le 6 c_1$, 
together with the effectivity requirement on $\eta$, force $\eta$ to be 
proportional to
$c_1$ (and so the model leads to no generations, because of $c_1^2=0$) as
one convinces oneself in the equivalent basis given by a section, the elliptic 
fibre (being equal to $c_1$) and the $E_8$ lattice. 

\section{Free ${\bf Z}_n$ symmetries}

\subsection{Modding the spaces}

Here we construct elliptically fibered Calabi-Yau threefolds $Z$ with sections, 
which admit an action $\iota$ of an discrete group $G$, such that the following 
requirements are met: 
\begin{itemize}
\item a.) $Z$ is smooth, 
\item b.) The action is free, i.e. the fixpoint set is empty, 
\item c.) The action leaves the holomorphic $(3,0)$-form invariant. 
\end{itemize}
We call $Z'=Z/G$, let $pr$ denote the projection. Note that $Z'$ is again an 
elliptic fibration (of $B$ type elliptic fibre) over a base $B'$. 
To get a $Z_n$ action we will start with the elliptic 
fibration with 
$n$ sections and use the algebraic $Z_n$ action from sect. \ref{spaces}, 
together with an in general non-free order $n$ action on the base. More 
precisely
what happens is this (let us put $n=2$, the case of our main example). 
As we consider an involution $\iota$ compatible with the fibration we
have an involution $\underline{\iota}$ already defined on $B$ alone with
$\underline{\iota} \cdot \pi=\pi\cdot \iota$ (note that $B'=B/\underline{G}$).
The group action $\iota$ 
maps both sections onto each other, their image downstairs in $Z'$ will be an
irreducible surface $\sigma$ (still isomorphic to $B$). Is this $\sigma$
a section in $Z'$ ? Now, from $pr^* \sigma \cdot pr^* f=2\;\, \sigma \cdot f$
where $f$ denotes the fibre downstairs (lying over a generic point $b'$ in the 
base $B'$
of $Z'$), one learns that it is actually only 
a {\em bi-section} because the left hand side is evaluated as $4=2+2$ as each
of the two sections $\sigma_i$, $i=1,2$, in the preimage of $\sigma$ intersects
each of the two fibers in $Z$ 'above' (lying over the preimages of $b'$ in $B$)
twice. So in general after the modding 
$n$ sections become one $n$-section. One looses therefore at least 
$n-1$ independent divisor classes in $H^{1,1}$.

In the following we will describe more closely the modding process: first
in section (4.1) for a Hirzebruch basis; here we find free involutions
for $F_0$ and $F_2$, so they will not lead to $3$ generations on 
$Z'$ as a 
corresponding $6$ generation model on $Z$ was found in the last section 
only over $F_1$; 
then in section (4.2) we find a free involution for the case 
$B_2=dP_9^B$ of $B$-type elliptic fibre.

\subsubsection{Toric descriptions}

In the following we will describe Calabi-Yau spaces $Z$ defined as hypersurfaces
in toric ambient spaces, with the three fiber types $A,B,C$. As we 
mentioned in the introduction, the 15 2-d toric reflexive polyhedra $\Delta_B$ 
from [\ref{koelman}] lead to smooth fibrations for all fibre types. The 
four dimensional polyhedra, which define the Calabi-Yau threefold  
following [\ref{batyrev}], are given by the convex hull of the points 
$\{[0,0,-1,0],[0,0,0,-1],[\nu^B,\nu^{(i)}])\}$, where the $\nu^B$ runs 
over the points in $\Delta_B$ and 
$\nu^{(1)}=(2,3)$,  
$\nu^{(2)}=(1,2)$ and  
$\nu^{(3)}=(1,1)$  for the $A,B,C$ fiber types. By the formulas 
derived in 
section 
\ref{cohomology}, we can express the cohomological information of $Z$ 
entirely in terms
of the base cohomology.  
  
\beqa
c_3(Z)&=&-2 h c_1(B)^2 \label{intersect1} \\
c_2(Z) J_E&=& k  c_2(B)+k({12\over k}-1) c_1(B)^2, \quad c_2  J_i= 12 k 
c_1(B) 
J_i  \\
J_E^3&=& k  c_1^2(B), \quad  J_E^2 J_i^B= k c_1(B) J_i^B, \quad J_E J_i J_k=k  
J_i 
J_k , 
\label{intersect3}
\eeqa
where the lefthand side is integrated over $Z$ and the righthand side over $B$.
$J_E$ is as cohomology element supported on the elliptic fibre, its dual 
homology 
element is the base, while the $J_i$ are as cohomology elements supported on 
curves in $B$, while the homology element is the dual curve in $B$ and 
the fiber over it.

In order to see whether requirements a.)-c.) are met we give a coordinate
description of $Z$, which turns out to be a straightforward generalization of 
the notion 
of a homogeneous polynomial in a projective space. We shall proceed in an 
example and 
consider as bases $F_n$ $n=0,1,2$. The polyhedron $\Delta_B$ contains the points 
$[1,0],[0,1],[-1,0],[-1,-n],[0,0]$ and hence $\Delta$ is given by the convex 
hull 
of points $\nu_i$ $i=0,\ldots 7$ in ${\bf R}^4$ equipped with the standard 
${\bf Z}$-basis for a lattice $\Lambda$ 

\begin{center}
\begin{tabular}{c|ccc|c}
[\ps 0,\ps 0,\ps 0,\ps 0] &  0 & 0 & -4 &$x_0$  \\ \hline
[-$n$,-1,\ps 1,\ps 2] & 1 & 0  & 0 & $s=x_1$\\ \hline
[\ps 0,\ps 1,\ps 1,\ps 2] &  1 & 0 & 0 & $t=x_2$ \\ \hline
[\ps 1,\ps 0,\ps 1,\ps 2] &  0 & 1 & 0 & $u=x_3$\\ \hline
[-1,\ps 0,\ps 1,\ps 2] & $-n$ & 1 & 0 & $v=x_4$ \\ \hline
[\ps 0,\ps 0,-1,\ps 0] & 0 &  0 & 1  & $x=x_5$\\ \hline
[\ps 0,\ps 0,\ps 0,-1] & 0 &  0 & 2  & $y=x_6$\\ \hline
[\ps 0,\ps 0,\ps 1,\ps 2] & $n-2$ & -2 & 1 & $z=x_7$ 
\end{tabular}
\end{center}
where we also indicated as columns the generators of linear relations among the 
$\nu_i$ 
$l^{(i)}$ spanning the Mori cone, which is dual to the K\"ahler cone. 
We can now write $Z$ as  hypersurface representation 
in coordinates $(s,t,v,u,x,y,z)$, one for each $\nu_i$, 
as
\beqa
x_0 y^2=
x_0 y\sum_{l=0}^2 x^l z^{2-l}\sum_{k=k^1_{min}}^{4-2l-k}v^k u^{4-2l-k} 
g_{d^1_{k,l,n}} +
x_0\sum_{l=0}^4 x^l z^{4-l}\sum_{k=k^2_{min}}^{8-2l-k}v^k u^{8-2l-k} 
f_{d^2_{k,l,n}} \ . \label{poly}
\eeqa
Here $d^1_{k,l,n}=(2-n)(2-l)+nk$ and $d^2_{k,l,n}=
(4-l)(2-n)+nk$ are the degrees of $f(s,t),g(s,t)$ in $s,t$ and 
the lower bounds on $k$ is such that this degree does not 
negative\footnote{$[x]$ 
means the next 
integer greater then $x$.} 
$k^1_{min}=\left[(2-n)(l-2)\over n\right]$ and $k^2_{min}=\left[(2-n)(l-4)\over 
n\right]$.
The form  [\ref{poly}] is manifestly invariant under 
the ${\bf C}^*$-actions $x_i\rightarrow \lambda_\mu^{l^{(\mu)}_i}x_i$,
$\lambda_\mu\in {\bf C}^*$, $\mu=1,2,3$ and therefore welldefined in the 
coordinate 
ring\footnote{Known as Batyrev-Cox coordinate ring.} of 
the $x_i$ ${\cal R}=\{C[x_1,\ldots,x_7]-{\cal SRI}\}/({\bf C}^*)^3$ where the 
Stanley-Reissner 
ideal ${\cal SRI}$ is generated by $x_0=0,x_1=x_2=0,x_3=x_4=0,x_5=x_6=x_7=0$. 
We may set
$x_0=1$ in the following. Note in particular
that for $n>2$ one has  $k^i_{min}>0$ and hence [\ref{poly}] becomes singular, 
with singularities that
can be directly read off e.g. an $A_3$ near $v=0$ for $n=3$ e.t.c. 
In general, the failure of transversality of the constraint has by Bertinis 
theorem only 
to be checked on the base locus, i.e. for the constraint $p=0$ restricted to 
hyperplanes 
defined by $\{x_i=0\}$, which also have to avoid of course the ${\cal SRI}$. 
Still 
this
is a formidable task in general, but for 3-d Calabi-Yau manifolds it is 
equivalent 
to reflexivity of $\Delta $ [\ref{batyrev}]. This fact has been used to check 
that the fibrations over bases from [\ref{koelman}] are indeed 
smooth and will be used to show that the member 
of the family admitting the $G$-action is smooth. Using the description of the 
cohomology by 
[\ref{batyrev}] we get $h_{1,1}=4(1)$ in all cases and $h_{2,1}=148(0)$ for 
$F_0$ and $F_1$ and $148(1)$ for $F_2$. Alternatively we can fix most of 
the automorphism group of the toric ambient space   
by writing the equation (\ref{poly}) into a more specialized form
\beqa
y^2=4 x^4+x^2 z^2 a(s,t,u,v)+x z^3 b(s,t,u,v)/3+z^4 c(s,t,u,v)/4, \label{poly2}
\eeqa
where the weight constraints on $a,b,c$ are as in (\ref{poly}). The part of the 
automorphism group, which is not fixed by this choice is the automorphism group 
of the base and an overall scaling. Then the discriminant of (\ref{poly2}) is
sufficiently simple to be analysed directly\footnote{$g_2={1\over 12}(a^2+12 c)$ 
and 
$g_3={1\over 216} (36 a c- 6 b^2- a^3)$}  
\beqa
\Delta=(9 a^4 c + 36 a b^2 c-72 a^2 c^2+144 c^3- a^3 b^2- 3b^4).\label{disc}
\eeqa  
Starting  from (\ref{poly2}) we may count the $h_{2,1}$ forms by 
enumerating perturbations modulo the remaining automorphism.
E.g. for $F_0$ $a,b,c$ are sections of line bundles corresponding to
polynomials of bi-degree 
$(4,4)$, $(6,6)$ and $(8,8)$ over $F_0={\bf P}^1\times {\bf P}^1$, 
which yields $h_{2,1}=5^2+7^2+9^2-3-3-1=148$ (cf. section (\ref{f0examples})).

In order that the holomorphic three-form is not projected out the total action 
of 
the
${\bf Z}_n$ on the cotangent space of $Z$ must be a subgroup of $SU(3)$. 
The condition is easily checked on the explicit expression of the holomorphic 
three-form $\Omega\sim {\prod_{i\neq k,l,m,p}x_i d x_k d x_l d x_m \over 
\partial_{x_p} p}$ from which we see that phase actions $x_i\rightarrow 
\mu^{w_i} 
x_i$ with 
$\mu^n=1$ and $\sum w_i=0\ {\rm mod }\ n$  leave $\Omega$ invariant. Hence 
${\bf Z}_2$ 
acting on the Calabi-Yau $Z$ by 
\beqa
(s,t,u,v,z,x,y)\rightarrow (-s,t,u,-v,-x,-y,z) \label{Z2action}
\eeqa
is an action with property c). To find the fix point set we must take into 
account 
the $\cal SRI$ and the ${\bf C}^*$ actions. If $n$ is even then 
$(0,t,u,0,0,0,z)$ 
is 
a fixed stratum in the ambient space, the $\cal SRI$ enforces 
$z\neq 0$, $t\neq 0$ and $u\neq 0$ and (\ref{poly}) restricted to the stratum 
is empty
($z^4u^8t^{4(2-n)}=0$ in contradiction with the $\cal SRI$) on $Z$. Similarly 
the 
other
fixed strata on which the action (\ref{Z2action}) can be undone by the ${\bf 
C}^*$-actions
have no intersection with $p=0$. On the other hand if $n$ is odd then 
by the ${\bf C}^*$ actions $(s,0,u,v,x,0,z)$ turn out to be the fixed stratum, 
which intersects $Z$ in particular for $n=1$ in a 
$\chi_{fix}=-16$ curve.

Hence to have a.) we shall restrict ourselves to the fix point free 
actions on $F_0$ and $F_2$ and consider transversality of the specialized 
polynomial which admits the ${\bf Z}_2$ involution. We may check transversality  
by showing that the polyhedron, which corresponds 
to the invariant monomials is again reflexive. The invariant polyhedron 
$\Delta'$ is constructed as in [\ref{kko}] by considering a sublattice 
$\Lambda'$ 
of index 
${1\over 2}$ in $\Lambda$. In the new ${\bf Z}$-basis the points transform to 
$\nu_i'=M_{i,j}\nu_j$ with 
\beqa
M=\left(\matrix{1&1&0&0\cr 0&1&1&0\cr 0&0&1&1 \cr 0&0&0&2}\right)
\eeqa
By constructing $\Delta'(F_0)$, $\Delta'(F_2)$ we see that they are 
indeed reflexive polyhedra and the  cohomology of the associated 
Calabi-Yau space is $\chi=-144$, $h_{1,1}=3$ and $h_{2,1}=75$, 
with one non-generic complex structure modulus 
for the $\Delta'(F_2)$ case\footnote{In 
accordance with the fix point formula 
$\chi^{new}=(\chi- \chi_{fix})/N + N \chi_{fix}$, we
find for $\Delta'(F_1)$: $\chi=-168$, $h_{1,1}=4$ and $h_{2,1}=88(9)$, i.e. 
nine non-generic complex structure moduli.}. Again in the $F_0$ case it
is simpler to count in (\ref{poly2}) the invariant monomials in $a,b,c$, 
which yields $13+24+41$. Note that each ${\rm SL}(2,{\bf C})$ is broken by 
(\ref{Z2action}) to the diagonal rescaling. Hence
$h_{2,1}=78-3=75$. The change in $(1,1)$ forms from $4$ to $3$ is due to the 
fact
that the $2$ sections become a $2$-section, as explained above. 
Note also that (\ref{disc}) remains generic since enough 
terms in $(a,b,c)$ survive the projection, hence $Z$ and so $Z'$ is smooth.
I.e. for these cases all requirements a.)-c.) are fulfilled.

Similarly we can obtain a free ${\bf Z}_3$
action on the $C$-fiber over ${\bf P}^2$. If we
introduce coordinates $(s,t,r)$ for the base and $(x,y,z)$ 
for the fibre as before, then the action
\beqa
(r,s,t,z,x,y)\rightarrow (\mu_3 r,\mu_3^2 s,t,\mu_3^2 x,\mu_3y,z) 
\label{Z3action}
\eeqa 
is a free ${\bf Z}_3$ which changes the Hodge numbers from 
$\chi=-216$, $h_{2,1}=112$, $h_{1,1}=4(2)$ to
$\chi=-72$, $h_{2,1}=38$ and $h_{1,1}=2$. 

{\em An afterthought}

Let us remark about some related points in constructing free involutions.
In two dimensions analogue there is a famous construction which 
is somewhat analogous,
the free Enriques involution
(killing the holomorphic two-from) on (certain) $K3$ . 
This gives as quotient again a surface
elliptically fibered over ${\bf P}^1$ of the
same Hodge numbers as $dP_9$ but has generically $10$ deformations 
and two {\em multiple}
fibers (so in particular does not have a section; if one performs 
logarithmic transformations on them one connects to the $dP_9$ surface), 
whose difference constitutes the canonical class which is two-torsion 
(after all, it vanishes 'above'), a second marked difference to $dP_9$.

Note that if we take in 
our construction of $B$ fibered Calabi-Yau (which are also $K3$ 
fibrations) just the $K3$ fibre part (i.e. 
neglect the overall base ${\bf P}^1$ and the part of the 
involution operating there) this 
leads to a $K3$ cover of an Enriques 
(the well-defindness of the operation restricts one then in the
deformation space of the $K3$ where the $B$ type fibration is kept 
manifest to a nine-dimensional space).

Note further that if one would take just the Calabi-Yau 
$(T^2\times K3_{Enr})/{\bf Z}_2$ 
and considered it as $K3$ fibered over the 
first factor\footnote{if one fibres 
over the second factor [\ref{DLOW}] one encounters difficulties with the 
effectiveness restrictions as the Enriques base then has a 
torsion canonical class}
then one gets a Calabi-Yau with a fundamental group 'including' 
${\bf Z}_2$ but this model has no section (although the generic $K3$ 
fibre has a section, over four points in the base 
one has the Enriques as fibre
which has no section but only a bi-section); 
as one would here have 'above'
an $N=2$ situation one has there no chiral matter.

\subsubsection{An example over $dP_9$}

Let us consider now the $B$ model fibration over $dP_9$, i.e. the Calabi-Yau 
space 
$Z=dP_9^B\times_{P^1} dP_9$ (the first fibre factor indicates the vertical
$dP_9$, the second one the horizontal base). We will choose for the base 
the $dP_9^B$ too, in contrast to the choice of $C'$ model we did in the 
generation search in section 3.8.
We want to supplement the involution $(y,x)\ra (-y,-x)$ in the fibre
by a second involution coming from the base. For this we `combine' (but see next 
footnote) 
the action $(s,t)\ra (-s,t)$ in the base ${\bf P}^1$ with 
the hyperelliptic involution in the fibre of the horizontal base $dP_9$.

So let $(x,y,z)$ be the coordinates of the fiber  of the vertical $dP_9$, 
$(s,t)$ the coordinates of the common base ${\bf P}^1$ and $(x',y',z')$ the
coordinates of the fiber of the horizontal ${\bf P}^1$. 
Let us recall the coordinate form 
\beqa
y^2+x^4+a_2(s,t)x^2z^2+b_3(s,t)xz^3+c_4(s,t)z^4=0 \label{dp9}
\eeqa 
of the vertical $dP_9^B$, with ${\cal R}=( C[x,y,z,s,t]-{\cal SRI} )/( {\bf 
C}^*)^2)$ where ${\cal SRI}$ is generated
by  $x=y=z=0,s=t=0$ and $l^{(1)}=(1,2,1,0,0),l^{(2)}=(0,0,-1,1,1)$,
shows that monomials within $a,b,c$ are invariant under the action; 
i.e. we are not forced to a singular locus like $b=0$.

This is easily enhanced to a fibre product structure, which is given
by the following complete intersection (we choose $B=dP_9^B$)
\beqa
y^2+x^4+a_2(s,t)x^2z^2+b_3(s,t)xz^3+c_4(s,t)z^4&=&0 \nonumber \\ 
y'^2+x'^4+{\tilde a}_2(s,t)x'^2z'^2+{\tilde b}_3(s,t)x'z'^3+{\tilde 
c}_4(s,t)z'^4&=&0  
\label{fiberproduct}
\eeqa 
The ${\bf C}^*$-actions are specified by 
\beqa
l^{(1)}&=&(1,2,\ps 1,0,0,0,0,\ps 0) \nonumber\\
l^{(2)}&=&(0,0,   -1,1,1,0,0,-1) \nonumber \\ 
l^{(3)}&=&(0,0,\ps 0,0,0,1,2,\ps 1)   
\label{cstar}
\eeqa

Then we consider the action
\beqa
(x,y,z,s,t,x',y',z')\rightarrow (-x,-y,z,-s,t,x',-y',z'), \label{z2dP9}
\eeqa
i.e. as before we supplement the involution in the fibre
by a second involution acting on the  base $dP_9$. 
Clearly this operation leaves the holomorphic $(3,0)$ form invariant
as $(-1)^4=+1$. 

\figinsert{fibreproduct}
{
The fibre product $Z=dP_9^{vert}\times_{P^1} dP_9^{hor}$, showing
the fixpoints of the action restricted to the factors. }
{2.5truein}{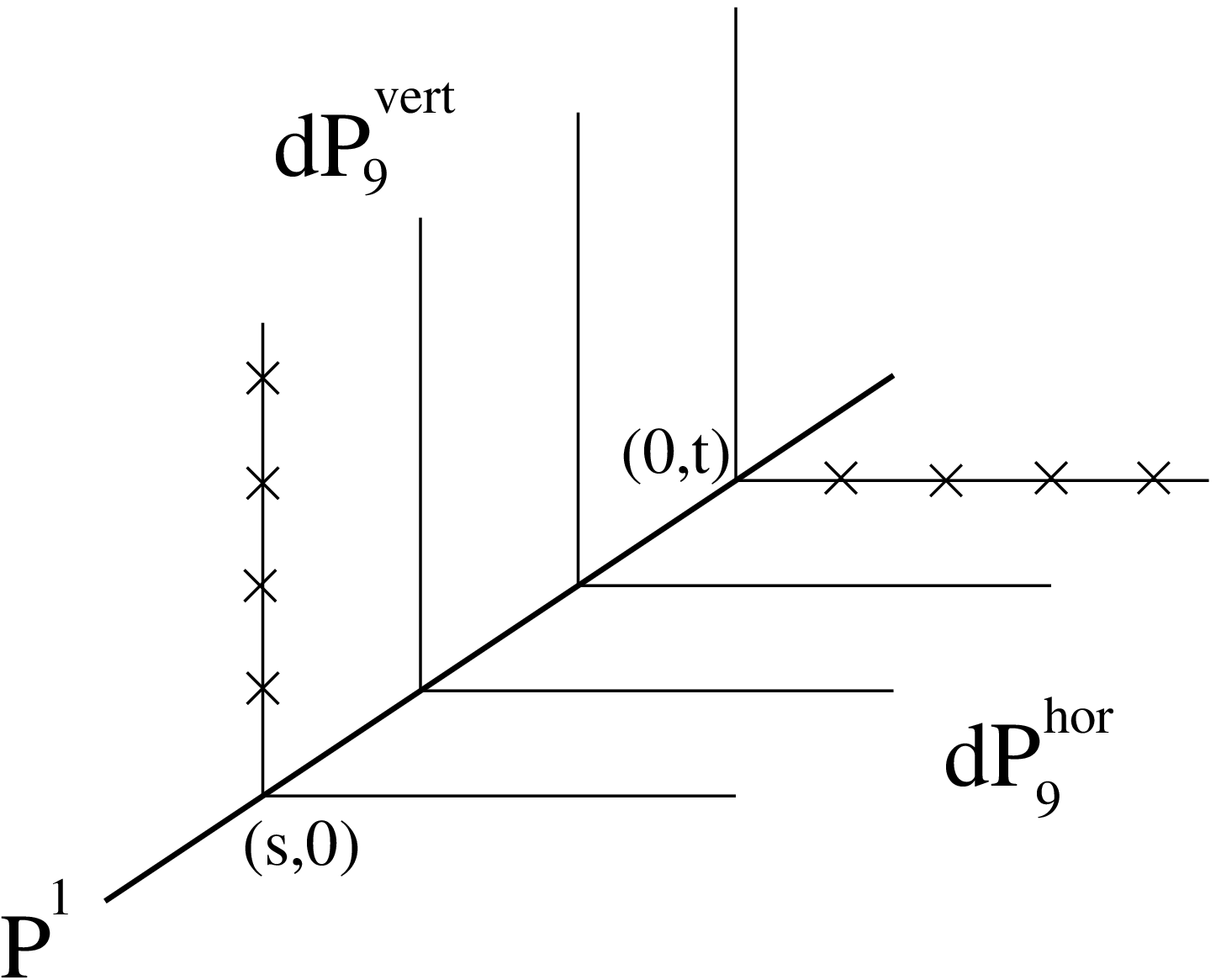} 

Let us look for possible fix loci by searching strata in the coordinates 
$(x,y,z,s,t,x',y',x')$. We start by distinguishing between two cases, which are 
necessarily
to have fixed strata, namely either $s\neq 0$ and  $t =0$ or $s=0$ and  $t\neq 
0$. 
In the first case the largest possible fixed strata are either 
$(x,0,z,s,0,0,0,z')$ 
or $(x,0,z,s,0,x',0,0)$, which both do not intersect (\ref{fiberproduct})
as the second equation forces either ($z'=0$ or $s=0$) or $x'=0$ all 
in the ${\cal SRI}$ and therefore excluded. In the second case the largest 
possible
strata are either $(x,0,0,0,t,x',0,z')$ or $(0,0,z,0,t,x',0,z')$, which for 
similar
reasons as above do not intersect with the first equation in 
(\ref{fiberproduct}).
If we look for the first case only in the vertical fibre then we see here  
$4$ fixpoints over one tip\footnote{Note here a special property of our 
construction of $dP_9$ in a nontrivial (weighted) ${\bf P}^2$ bundle as opposed 
to 
the usual
${\tiny \left[\begin{array}{c|c}P^2&3\\P^1&1\end{array}\right]}$: In that case 
one 
would really
have over both fixpoints in the base an hyperelliptic involution with $4$ fixed 
points. In that sense our
involution is not just a naive combination of the ${\bf P}^1$ involution and the 
hyperelliptic 
involution.}  of ${\bf P}^1$ at $t=0$. Likewise for the second case, 
if we look only in the horizontal fibre, we see here $4$ fixpoints over
the other tip of ${\bf P}^1$ $s=0$. The fact that, on the other hand, 
at $t=0$ ($s=0$) in the ${\bf P}^1$ the action in the horizontal (vertical) 
fibre is the shift $(y',x')\rightarrow (-y',-x')$ ($(y,x)\rightarrow (-y,-x)$) 
is a more conceptual way to see that the total action is actually free.

The number of deformations which is $h^{2,1}(Z)=2(3+4+5)-3-1-1=19$ on the 
covering 
drops by the modding to $h^{2,1}(Z')=2(2+2+3)-1-1-1=11$.
Likewise by the Lefschetz fixpoint formula one finds for the number of surviving 
$(1,1)$-forms $ n_{1,1}^+=11$ as 
$0=2+2(n_{1,1}^+-n_{1,1}^-)-[2+2(n_{2,1}^+-n_{2,1}^-)]=
2+2(2 n_{1,1}^+-19)-[2+2(11-8)]$.  
What concerns the smoothness of $Z$ note that for the individual factors one 
can easily check using eq. (\ref{disc}) that the invariant
monomials, just counted, are generic enough to keep the space smooth.
So, as the Euler number remains zero, we mod from a $(19,19)$ Calabi-Yau $Z$ to 
an 
$(11,11)$
Calabi-Yau $Z'$. Note that again our conditions a.)-c.) are fulfilled.

\subsection{Modding the bundles}

In the 'old' case of using the tangent bundle, obviously the bundle was carried
along when just modding out the space: one knows then that the bundle 'above'
on the Calabi-Yau $Z$ is a pullback of the tangent bundle 'downstairs' of the
modded Calabi-Yau $Z'$ and it follows, in the case of modding by an involution 
$\iota$ 
(the case of interest here for us),  that the Euler number relevant for the net
generation number ($=$ Euler$/2$) is accordingly reduced, 
i.e. $\int_Z T(Z)=2\int_{Z'}T(Z')$. To 'mod' the bundle in the more general
case one has just to assume that one really starts from a bundle on $Z'$ where
one later breaks the gauge group by turning on Wilson lines using 
$\pi_1(Z')={\bf Z}_2$. On the other hand to compute the generation number one
has to work 'upstairs' as $Z'$ does not have a section but only a bi-section 
(left over from the two sections of $Z$ as discussed above)
and so one can not use the spectral cover method, as developed so far, directly 
on $Z'$: remember that the crucial translation from bundle data to geometrical
data occurred when a bundle (decomposing fibrewise into a sum of line bundles)
is described fibrewise by a collection of points, for which a choice of 
reference
point $p$ (the "zero" in the group law) has to be made (to represent a 
degree zero
line bundle by the divisor $Q_i-p$); one then needed, for this collection of 
points to fit together to the branched $n$-fold covering $C$ of $B$, the 
{\em existence of a global reference point}, i.e. a section. So we will 
actually lift back the bundle $V'$ we have downstairs on $Z'$ 
(whose generation number we want to compute as $3$) to a bundle $V$ on $Z$
which should then, because of $\int_Z V=2\int_{Z'} V'$, have $6$ generations
(and is then 'moddable' by construction).

Now, how can one be sure when one constructed a bundle above with $6$ 
generations that it actually was such a pull-back from $Z'$ ? 
Clearly for this $V$ should be $\iota$ invariant.
For this let us make the following consideration\footnote{Note that we consider
fibration compatible involutions, which are thus already defined on the base
$B$ alone and map a fiber over a base point to another fibre.}. 
{\em First}, when one just has a bundle on $Z$, where $Z$ is
a $B$ type elliptically fibered Calabi-Yau, one can $V$ translate into 
geometrical data in two ways: 
namely one can analyze $V$ with respect to both global reference
points (i.e. sections) giving the alternative descriptions 
$C_i=n\sigma_i+\eta_i$
for $i=1$ and $2$. But as it is both times actually the same bundle 
all invariants associated with $V$, such as the Chern classes, must coincide. 
This will be the case if both $\eta$'s and $\lambda$'s coincide (this
is just a sufficient condition).

Now, {\em secondly}, 
let us assume that we are actually analyzing a bundle $V$ on $Z$
which is a pull-back from a bundle $V'$ on $Z'$. Again we can analyze $V$ 
relative to $\sigma_1$. But now we have to make explicit the condition of
$\iota$-invariance, i.e. (as the spectral cover of $\iota^*V$ with 
respect to the $\sigma_2$ would be\footnote{the transport by $\iota$ of
line bundles over an elliptic fibre, which is a map between $Pic_0$'s,
is, when the $Pic_0$'s are again identified with the elliptic fibre,
again the map $\iota$} $\iota C_1$)
we have to move $C_1$ by $\iota$ (of course, as the geometrical
information encoded in a spectral cover as $C_1$ is defined only relative to
the global reference point $\sigma_1$, one has for the correct interpretation
of $\iota (C_1)$ to read it relative to $\iota (\sigma_1)=\sigma_2$) and get
a spectral cover $\iota (C_1)$ over $\sigma_2$ of class $[\iota 
(C_1)]=n\sigma_2+
\iota(\eta_1)$; so, from what was said in the preconsideration and as 
we require now $\iota^*V=V$, one finds the $\iota$-invariance 
condition fulfilled for $\iota(\eta_1)=\eta_1$ (and $c_1(B_2)$ should 
be fix, too; similar considerations pertain to $\gamma$; again we are
presenting sufficient conditions, but note that in our case ($B=F_0$ or $F_2$)
the $\iota$ operation induced on $H^{1,1}(B)$ will be trivial anyway).

Now, in the two actual examples over $F_0$ and $F_2$, where we could mod the
Calabi-Yau, one has a basis of $H^{1,1}(B)$ consisting in the classes $b$ and
$f$, the base (of negative self-intersection, say) and the fibre class. Now, 
in Dolbeault cohomology the relevant representatives of the top $(1,1)$ form 
of these
two ${\bf P}_1$ are given in local coordinates by, say, $dzd\bar{z}$; but these
classes are invariant under the phase rotations (actually $(-1)$'s) 
on the ${\bf P}_1$. So that condition does not represent a restriction.

Note that this {\em could} be a true restriction. For example,
in case of a $dP_9$ base, the involution in the base (which can be non-free)
could have, say, $8$ or $4$ fix points  (in the case\footnote{Cf. the 
footnote 22 the case of $dP_9^{C'}$.} 
of $dP_9^{C'}$ or $dP_9^B$) if we combine the $(s,t)\ra (-s,t)$ operation in the 
base ${\bf P}_1$ with the hyperelliptic involution.  By Lefschetz we see
that, say, $8=2+n_+-n_-$ with $n_{\pm}$ the $\pm 1$ eigenclasses of the induced
$\iota$ operation on $H^{1,1}(B)$; from $n_++n_-=10$ one gets that $n_-=2$ resp.
$4$ classes for the $4$ fixed point case are actually projected out. By contrast
in our cases of Hirzebruch bases, where our involution not only respects the
elliptic fibration of the Calabi-Yau but also the internal 
${\bf P}_1$ fibration of the base, our operation there again 'factorizes'
into operations (like $(s,t)\ra (-s,t)$) of both the ${\bf P}_1$'s, so
that from the two individual fixed points on each of them one gets
four overall fixed points in $B$ and so, of course obvious anyway, from
Lefschetz in that case that $n_-=0$, i.e. no real restriction results.

\section{Genuine $B$ type bundles}

Up to now the influence of the choice of a $B$ type elliptic fibered base 
Calabi-Yau
$Z$ had a rather restricted impact on the general set-up. Essentially the 
influence
of this alteration was restricted to the change in $c_2(Z)$ which had its 
consequences
for the upper bound of $\eta$ and a possibility to build the 'other' bundle 
'over'
$\sigma_2$. But there is an even more interesting twist in the story which we 
have 
not mentioned before as we wanted to keep the different steps of complication of 
the 
building of bundles disentangled as far as possible. The new freedom we are now 
speaking about stems from the fact that we have more divisors in the game and so
a greater chance to build up a line bundle on the spectral cover. Remember that 
in
the push-forward construction of $V$ from $L$ the class of $L$ was constrained 
by
the requirement $c_1(V)=0$; this determined $c_1(L)$ up to a class in 
$ker\, \pi_*:H^{1,1}(C)\ra H^{1,1}(B)$ killed by the
push-forward\footnote{we neglect here the continuous moduli from $H^{1,0}(C)$}.
In the $A$ model one has two obviuos sources of divisors, consisting of either
pull-back's from $B$ or the section (restricted to $C$); then the combination
in $ker\, \pi_*$ was $\gamma=\lambda(n\sigma-\eta+nc_1)$. Now, with a second 
section,
a new option arises:
\beqa
\delta=\mu(\sigma_1+c_1-\sigma_2)
\eeqa
One has $\gamma\cdot \delta=0$, $\pi_*\delta^2=-2\mu^2 \eta c_1$ and finds that the 
general combination of $\gamma+\delta$ can be used in building up $L$ with $\mu$
integral and $\lambda$ restrictions unchanged. 
The influence in cohomological data is (here a major computation is to be 
carried out 
which for the $A$ fibre is outlined in the appendix of [\ref{C}])
\beqa
c_2(V)&=&\eta\sigma_1 - \eta \delta -\frac{n^3-n}{24}c_1^2+
\frac{1}{2}(\lambda^2-\frac{1}{4})n\eta (\eta-nc_1)+\mu^2\eta c_1\nonumber\\
\frac{1}{2}c_3(V)&=&\lambda \eta (\eta-nc_1)-\frac{3}{2}\mu\eta 
c_1(\sigma_1-\sigma_2)
=\lambda \eta (\eta-nc_1)
\eeqa
where the generation number is unchanged because $\int_Zc_3(V)=0$ as both 
$\sigma_i$
are sections leading after integration over the fibre to an integral over $B$ 
times
$(1-1)$. Of course that $\delta$ doesn't contribute is not an accident: the 
matter
computation, described in [\ref{C}] too, shows that the matter is localized on 
the
curve $A=C\cdot \sigma_1\subset \sigma_1(B)$; as $V|_{\sigma_1(B)}=\pi_*L$ the 
computation of $h^0(Z,V)-h^1(Z,V)$ is reduced to 
$h^1(A,L|_A\otimes K_A^{1/2})-h^0(A,L|_A\otimes K_A^{1/2})=-c_1(L|_A)=
-deg\, 
\gamma|_A=-\gamma\cdot\sigma_1=\lambda\eta\sigma_1=\lambda\eta(\eta-nc_1)$ 
where in the last equality the intersection number was evaluated in 
$B\cong \sigma_1(B)$ instead of $C$. But the corresponding consideration for 
$\delta$
shows that is doesn't contribute as $\delta \cdot \sigma_1=0$.

So one must still search for $6$ generations with
the pure $\gamma$ formula, i.e. this generalization is not capable
of generating chiral matter by itself, and allows one (within the mentioned $\eta$-bounds) 
not, even when combined with the
ordinary $\gamma$ class, to find $6$ generation models over a Hirzebruch surface 
where one can 'mod'.

\section{A further bundle parameter}
 
As the two requirements in our search, $6$ net generations and a moddable 
Calabi-Yau, have lead us so far to mutually exclusive examples (of bases 
$F_1$ respectively $F_0$, $F_2$ and $dP_9^B$), we have still to broaden our arsenal
of constructions. For this we will put the burden on the generation number
search and will use a second method to get chiral matter besides turning on
$\gamma$.

Recall that in section (3.1) we explained the idea of the spectral cover
construction to trade in the $SU(n)$ bundle $V$ over $Z$ for a line bundle
$L$ over a $n$-fold branched covering $C$ of $B$: one had
\beqa
V=p_*(p_C^*L\otimes {\cal P})
\eeqa
with $p:Z\times_B C\ra Z$ and ${\cal P}$ the
(global) Poincare line bundle over $Z\times Z$ (restricted to $Z\times_B C$).
We remarked there that the trade occurs 'essentially' over $B$, i.e.
on a {\em surface} level, which referred
to the fact that we got the line bundle ${\cal L}$ over $Z\times_B C$, which is
needed for the construction $V=p_*({\cal L}\otimes {\cal P})$, as a pull-back
{}From a line bundle $L$ over $C$. 

Actually the construction is naturally slightly generalized leading to
a dependence on a further parameter\footnote{a possibility already
mentioned in [\ref{BJPS}]}  [\ref{C}] when one takes 
into
account the full capability of constructing line bundles ${\cal L}$ over the
{\em three-fold} $Z\times_B C$: 
$Z\x _B C$ will have a set ${\cal S}$ of
isolated singularities (generically ordinary double points) 
when in the base the discriminant referring to
the $Z$-direction meets the branch locus referring to the $C$ direction. 
I.e. they are lying over points in the base $B$ where
the branch divisor $r=K_C-\pi^*K_B$ of $\pi :C \ra B$, resp. its image in $B$
$\pi_* r=n(2\eta -(n-1)c_1)$, meets the discriminant $12c_1$
(cf. the case of an elliptic  $K3$ over ${\bf P}^1$) of $\pi: Z\ra B$.
Their number is (as detected in $B$)
$|{\cal S}|=12c_1n(2\eta -(n-1)c_1)$.
One has a resolution $\nu :Y\ra Z\times _B C$ of the isolated
singularities with $E$ the exceptional divisor.
Each of its $|{\cal S}|$ components
is a divisor $D={\bf P}^1\x {\bf P}^1$ 
of triple self-intersection $D^3=(-1,-1)^2=+2$.

The resolution $Y\ra Z\x _B C$ leads to the possibility to formulate
the whole construction on $Y$ and twist
there by the line bundle
corresponding to a multiple $l\in {\bf Z}$ of the exceptional divisor $E$.
So one arrives at the 
description $V=p_* \nu_* {\cal L}$ where
${\cal L}=\nu^* p_C^*L \otimes \nu^* {\cal P}\otimes {\cal O}_Y(lE)$.
Including this twist one gets for the generations in total
\beqa
\frac{1}{2}c_3(V)=\lambda \eta (\eta -n c_1)+
\frac{l(l-1)(2l-1)}{6}|{\cal S}|\nonumber
\eeqa
Note that the additional contribution coming from $c_1({\cal O}_Y(lE))$
is even componentwise
integral as for example (for $l\ge 0$)
$\frac{l(l-1)(2l-1)}{6}=\sum_{i=1}^{l-1}i^2$.

Now what concerns the application in our search for $6$ generation models over 
$F_0$ or $F_2$, fate wants it that numerically it is still not possible to 
reach the number $\pm 6$, essentially because $|{\cal S}|=12c_1n(2\eta 
-(n-1)c_1)=
120c_1(\eta-2c_1)$ turns out to be a somewhat unpleasant large number which is
unsuitable to be tuned to $\pm 6$ even with the combined effort of the 
$\lambda \eta (\eta -5 c_1)$ term ($\lambda \in \frac{1}{2}+{\bf Z}$!).

But there is actually an easy way out. The exceptional divisor $E$ decomposes
into $|{\cal S}|$ components and we are not forced to buy its effect in the 
total 
collection. I.e., we do not have just a second discrete parameter 
$l$ besides $\lambda$, but this $l$ is actually a vector $(l_i)_{i\in {\cal S}}$ 
of $|{\cal S}|$ components
of integral numbers! The only thing we could be forced to is to take an 
$G$-orbit 
(where $G$ is the freely acting group by which finally we want to mod out; 
its operates on the set ${\cal S}$ of singularities) to secure 'moddability'.

But as in our case $G$ is just ${\bf Z}_2$, generated by the involution $\iota$,
this forces us at most to turn on two individual $l_i$ in parallel. When we take 
their common value to be $2$ (so that $\frac{l(l-1)(2l-1)}{6}=1$) their 
contribution
in the generation number is only $1\cdot 2=2$. But now it is easy, by turning on
enough $l_i$'s, to reach $6$.

\vskip 1cm

G.C. thanks E. Witten for pointing out the possibility of a construction
similar to the one considered here.

\section*{References}
\begin{enumerate}

\item
\label{W}
E. Witten, {\em New issues in manifolds of $SU(3)$ holonomy}, 
Nucl. Phys. {\bf B268} (1986) 79.

\item
\label{FMW}
R. Friedman, J. Morgan and E. Witten, {\em Vector bundles and $F$-theory},
Comm. Math. Phys. {\bf 187} (1997) 679, hep-th/9701162.

\item
\label{C}
G. Curio, {\em Chiral matter and transitions in heterotic string models},
Phys. Lett. {\bf B435} (1998) 39, hep-th/9803224.

\item
\label{koelman}
R. Jan Koelman, {\em A criterion for the ideal of a projectively embedded toric 
surface to be 
generated by quadrics}, Beitr\"age  Algebra Geom. 34 (1993), no. 1, 57--62.

\item
\label{Wsy}
E.Witten, {\em Symmetry breaking patterns in superstring models}, Nucl. Phys.
{\bf B258} (1985) 75.

\item
\label{KLM}
A.Klemm, W.Lerche and P.Mayr, 
{\em K3--Fibrations and Heterotic-Type II String Duality},
Phys. Lett. {\bf B357} (1995) 313, hep-th/9506112.

\item
\label{AFIU}
G. Aldazabal, A. Font, L.E. Ibanez and A.M. Uranga, 
{\em New Branches of String Compactifications and their F-Theory Duals},
Nucl. Phys. {\bf B492} (1997) 119, hep-th/9607121.

\item
\label{KMV}
A. Klemm, P. Mayr and C. Vafa, {\em  BPS States of Exceptional 
Non-Critical Strings}, Proceedings of the conference "Advanced 
Quantum  Field Theory'' (in memory of Claude Itzykson), hep-th/9607139.

\item
\label{KLRY}
A. Klemm, B. Lian, S.-S. Roan and S.-T. Yau, 
{\em Calabi-Yau fourfolds for M- and F-Theory compactification}, 
Nucl. Phys. {\bf B518} (1998) 515, hep-th/9701023.

\item
\label{BKMT}
P. Berglund, A. Klemm, P. Mayr and S. Theisen, 
{\em On Type IIB Vacua With Varying Coupling Constant}, hep-th/9805189.

\item
\label{KM}
A.\ Klemm, P.\ Mayr, {\em Strong Coupling Singularities and 
Non-abelian Gauge Symmetries in $N=2$ String Theory},
Nucl.Phys. B469 (1996) 37-50

\item
\label{KPM}
S.\ Katz, R.\ Plesser, D.\ Morisson {\em Enhanced Gauge Symmetry in 
Type II String Theory}, Nucl.Phys. B469 (1996) 37-50

\item
\label{MVII}
D.\ Morrison, C.\ Vafa, {\em 
Compactifications of F-Theory on Calabi--Yau Threefolds -- II},
Nucl.Phys. B476 (1996) 437-469 

\item
\label{reid}
M.\ Reid {\em Canonical $3$-folds} 
Journ\'ees de G\`eometrie Alg\`ebrique d'Angers, 
Juillet 1979/Algebraic Geometry, Angers, 1979, 273--310, 
Sijthoff \& Noordhoff, Alphen aan den Rijn

\item
\label{batyrev}
V.\ Batyrev, {\em Dual polyhedra and mirror symmetry for Calabi-Yau 
hypersurfaces 
in toric varieties}, J. Algebraic Geom. 3 (1994), no. 3, 493--535

\item
\label{kko}
S. Kachru, A. Klemm and Y. Oz {\em 
Calabi-Yau Duals for CHL Strings}, Nucl.Phys. B521 (1998) 58-70

\item
\label{DLOW}
R. Donagi, A. Lukas, B.A. Ovrut and D. Waldram, {\em Holomorphic 
vector bundles and non-perturbative vacua in M-theory}, hep-th/9901009.

\item
\label{A}
B. Andreas, {\em On vector bundles and chiral matter in $N=1$ heterotic
compactifications}, JHEP {\bf 01} (1999) 011, hep-th/9802202.  

\item
\label{BM}
P. Berglund and P. Mayr, {\em Heterotic String/F-theory Duality 
from Mirror Symmetry},
hep-th/9811217.

\item
\label{CD}
G. Curio and R. Donagi, Nucl. Phys. {\bf B518} (1998) 603, hep-th/9801057.

\item
\label{R}
G. Rajesh, {\em Toric Geometry and $F$-theory/heterotic duality in four
dimensions}, to appear in JHEP, hep-th/9811240.

\item
\label{BJPS}
M. Bershadsky, A. Johansen, T. Pantev and V. Sadov, 
{\em On Four-Dimensional
Compactifications of $F$-Theory}, Nucl. Phys. {\bf B505} (1997) 165, 
hepth/9701165.

\end{enumerate}
\end{document}